\documentclass[12pt]{article}
\usepackage{graphicx}
\usepackage{epsfig}
\usepackage{epstopdf}
\DeclareGraphicsExtensions{.pdf,.eps,.png,.jpg,.mps}

\def\lsim{\mathrel{\rlap{\lower3pt\hbox{\hskip0pt$\sim$}}
     \raise1pt\hbox{$<$}}}         
\def\gsim{\mathrel{\rlap{\lower4pt\hbox{\hskip1pt$\sim$}}
     \raise1pt\hbox{$>$}}}         

\usepackage{amsmath}
\usepackage{amsfonts}

\begin{document}
\begin{titlepage}

\centerline{\Large \bf 4-Factor Model for Overnight Returns}
\medskip

\centerline{Zura Kakushadze$^\S$$^\dag$$^\ddag$\footnote{\, Email: \tt zura@quantigic.com}}
\bigskip

\centerline{\em $^\S$ Quantigic$^\circledR$ Solutions LLC}
\centerline{\em 1127 High Ridge Road \#135, Stamford, CT 06905\,\,\footnote{\, DISCLAIMER: This address is used by the corresponding author for no
purpose other than to indicate his professional affiliation as is customary in
publications. In particular, the contents of this paper
are not intended as an investment, legal, tax or any other such advice,
and in no way represent views of Quantigic$^\circledR$ Solutions LLC,
the website \underline{www.quantigic.com} or any of their other affiliates.
}}
\centerline{\em $^\dag$ Department of Physics, University of Connecticut}
\centerline{\em 1 University Place, Stamford, CT 06901}
\centerline{\em $^\ddag$ Free University of Tbilisi, Business School \& School of Physics}
\centerline{\em 240, David Agmashenebeli Alley, Tbilisi, 0159, Georgia}
\medskip
\centerline{(October 19, 2014; revised June 4, 2015)}

\bigskip
\medskip

\begin{abstract}
{}We propose a 4-factor model for overnight returns and give explicit definitions of our 4 factors. Long horizon fundamental factors such as value and growth lack predictive power for overnight (or similar short horizon) returns and are not included. All 4 factors are constructed based on intraday price and volume data and are analogous to size (price), volatility, momentum and liquidity (volume). Historical regressions {\em a la} Fama and MacBeth (1973) suggest that our 4 factors have sizable serial t-statistic and appear to be relevant predictors for overnight returns. We check this by using our 4-factor model in an explicit intraday mean-reversion alpha.
\end{abstract}
\medskip

\end{titlepage}

\newpage

\section{Introduction}

{}Most quantitative trading nowadays is done on short horizons, ranging from a day (or a few days) down to micro- and even nanoseconds. Most commercially available multi-factor risk models, majority of whose users are institutions with long holding horizons such as mutual funds and pension funds, contain longer horizon risk factors unsuitable for quantitative trading strategies (Kakushadze and Liew, 2015). Factor models with fewer factors such as Size (SMB), Value (HML) and Momentum (WML), including the 3-factor model by Fama and French (1993), are also geared toward explaining longer horizon returns.

{}In this note, to fill the apparent vacuum in the literature, we propose a 4-factor model for overnight returns.\footnote{\, As we discuss below, we suspect that this model also ought to do well for intraday returns.} For such short horizons longer horizon factors such as value and growth are not relevant (Kakushadze and Liew, 2015). The question then is, what is relevant and how to construct such factors?

{}The fundamental principle we anchor on is what we refer to as ``Horizon Decoupling Principle". In a nutshell, what happens at time horizon $T_1$ is not affected by what happens at time horizon $T_2$, if $T_1$ and $T_2$ are vastly different.\footnote{\, By ``time horizon" we mean the relevant time scales. {\em E.g.}, the time horizon for a daily close-to-close return is 1 day.} In terms of returns, this decoupling can be restated as the returns for long-term horizons $T_1$ being essentially uncorrelated with the returns for short-term horizons $T_2$.

{}The Horizon Decoupling Principle greatly simplifies searching for relevant risk factors for overnight returns. We can immediately discard any factors based on longer horizon quantities such as value and growth ((Basu, 1977), (Fama and French, 1992, 1993), (Lakonishok {\em et al}, 1994), (Asness and Stevens, 1995), (Haugen, 1995), (Liew and Vassalou, 2000)), which change essentially quarterly. However, even for other factors such as size (Banz, 1981), momentum ((Jegadeesh and Titman, 1993), (Asness, 1995)), liquidity ((Scholes and Williams, 1977), (Asness, {\em et al}, 2001), (Pastor and Stambaugh, 2003), Anson (2013/14)) and volatility (Ang {\em et al}, 2006), care is needed in defining them. Let us take, {\em e.g.}, size. Normally, it is defined as a logarithm of market cap.\footnote{\, Typically, appropriately normalized cross-sectionally -- see below.} However, market cap is a product of price and shares outstanding. While price changes daily, shares outstanding on average change much less frequently and have little predictive power for overnight returns. This forces the size factor for overnight returns to be defined via a logarithm of {\em price}. We discuss this and other points in relation to other risk factors in more detail below.

{}What we end up with is a 4-factor model. All 4 factors are constructed based on intraday price and volume data and are analogous to size (price), volatility, momentum and liquidity (volume). Historical regressions {\em a la} Fama and MacBeth (1973) suggest that our 4 factors have sizable serial t-statistic and appear to be relevant predictors for overnight returns.\footnote{\, Typically, one would also include market beta as another factor. Here we use a unit vector, {\em i.e.}, the intercept instead, albeit one can use non-unit beta if desired. We discuss this point in more detail in Section 2.}

{}The remainder of this note is organized as follows. In Section \ref{sec2} we discuss relevant (to overnight returns) risk factors and give their explicit definitions. Section 3 contains concluding remarks. Tables contain results of our historical regressions, which we discuss in Section \ref{sec2}. Some legalese is included at the end of Section \ref{sec3}.

\section{Four Factors}\label{sec2}

{}Since, according to the Horizon Decoupling Principle, we should not use any long horizon factors, barring any betas (see below), we are left with 4 candidate factors: size, momentum, liquidity and volatility. Let us discuss them one-by-one. Some notations first. $P_i$, $i=1,\dots,N$ is the stock price for the stock labeled by $i$, where $N$ is the number of stocks in our universe. In actuality, the price for each stock is a time-series: $P_{is}$, $s=0,1,\dots,M$, where the index $s$ labels trading dates, with $s=0$ corresponding to the most recent date in the time series. If no confusion arises, we will sometimes omit the index $s$. We will use superscript $O$ (unadjusted open price), $H$ (unadjusted intraday high price), $L$ (unadjusted intraday low price), $C$ (unadjusted close price), $AO$ (open price fully adjusted for splits and dividends), and $AC$ (close price fully adjusted for splits and dividends), so, {\em e.g.}, $P^C_{is}$ is the unadjusted close price. $V_{is}$ is the unadjusted daily volume (in shares, not dollars). Also, we define the overnight return as the close-to-next-open return:
\begin{equation}
 R_{is} \equiv \ln\left({P^{AO}_{is} / P^{AC}_{i,s+1}}\right)
\end{equation}
Note that both prices in this definition are fully adjusted.

\subsection{Factor Model Structure}

{}We wish to construct a factor model of the form
\begin{equation}\label{reg}
 R_{is} \sim \sum_{A = 1}^K \beta_{iAs}~ f_{As} + \varepsilon_{is}
\end{equation}
Here $K$ is the number of risk factors, $f_{As}$ are the $K$ factor returns, $\varepsilon_{is}$ are the residuals, and $\beta_{iAs}$ are the factor betas. Note that we do not have an intercept in (\ref{reg}) because we include the intercept in $\beta_{iAs}$, {\em i.e.}, for a given date $s$, the $N\times K$ matrix $\beta_{iAs}$ contains a column equal the unit $N$-vector. This will be the first column in $\beta_{iAs}$. We will denote each such column by $\beta^{\rm{\scriptstyle{\{name\}}}}_{is}$ (without the index $A$), where \{name\} stands for the name by which we refer to the corresponding factor: ``int" for the intercept, ``prc" for price (analog of size), ``mom" for momentum, ``hlv" for intraday high and low based volatility, and ``vol" for volume (analog of liquidity). Here we wish to construct all factors without using any long horizon fundamental data or intraday tick data, only daily open, high, low, (adjusted) close and volume.

\subsection{Intercept (``beta")}\label{sub.beta}

{}The intercept plays the role of ``market beta":
\begin{equation}
 \beta^{\rm{\scriptstyle{int}}}_{is} \equiv 1
\end{equation}
Any other non-unit beta typically is too noisy, especially for such short horizons. To remove the noise, one would have to compute correlations of individual stock returns with the market return based on time series looking back many days, which would defy the purpose here as we wish to construct short horizon factors. Neutrality w.r.t. $\beta^{\rm{\scriptstyle{int}}}_{is}$ is simply dollar neutrality, which is approximate market neutrality.

\subsection{Price (``Size")}

{}Size is essentially a logarithm of market cap (before any cross-sectional normalization -- see below). However, market cap $C_i$ is a product of price $P_i$ and shares outstanding $S_i$. While price changes daily, shares outstanding $S_i$ on average change much less frequently and have little predictive power for overnight returns.\footnote{\, We have run the same historical regressions as for the factors we discuss below, with $S_i$ and $\ln(S_i)$ as risk factors, and confirmed that shares outstanding do not add value.} Since $\ln(C_i) = \ln(P_i~S_i) = \ln(P_i) + \ln(S_i)$, and since $\ln(S_i)$ does not add value, we focus on $\ln(P_i)$. To be precise, as the analog of the size factor beta for overnight returns, we will take
\begin{equation}\label{size}
 \beta^{\rm{\scriptstyle{prc}}}_{is} \equiv \ln\left(P^{AC}_{i, s+1}\right)
\end{equation}
{\em I.e.}, on date $s$ we use the previous day's adjusted close. More precisely, in historical regressions we will use two versions of this factor beta, one defined by (\ref{size}), and another one via
\begin{equation}\label{size1}
 {\widetilde \beta}^{\rm{\scriptstyle{prc}}}_{is} \equiv \ln\left(P^{C}_{i, s+1}\right)
\end{equation}
{\em I.e.}, on date $s$ we use the previous day's unadjusted close. It is more natural to use (\ref{size}) and this is what should be used in any practical application assuming the splits and dividends are known on the ex-date before the computation is performed. However, to preemptively pacify any concerns that the regressions we discuss below somehow ``look" into the future by using adjusted previous close, we use both versions. Not surprisingly, (\ref{size}) works better. However, (\ref{size1}) also works well. The truth lies in between, closer to the results corresponding to (\ref{size}).\footnote{\, Assuming the splits and dividends data is known with 100\% certainty on the ex-date before the computation, the regression results corresponding to (\ref{size}) are 100\% out-of-sample. However, in practice data providers sometimes can miss dividends for some stocks and fix the adjusted close retroactively, so in this sense the results based on (\ref{size}) may not be 100\% out-of-sample. While (\ref{size1}) is an overkill, it is 100\% out-of-sample as it relates to adjustments as none are applied.}

\subsection{Momentum}

{}There are various ways of defining the momentum factor beta. {\em E.g.}, one can take a $d$-day moving average of close-to-close returns, {\em etc.} Generally, the returns from the dates further and further into the past have lower and lower correlations with our overnight returns. Here we define the momentum beta as follows:
\begin{equation}\label{mom}
 \beta^{\rm{\scriptstyle{mom}}}_{is} \equiv \ln\left(P^{C}_{i, s+1} / P^{O}_{i, s+1}\right)
\end{equation}
{\em I.e.}, this is simply the previous day's open-to-close return.

\subsection{Intraday Volatility}

{}There are various ways of defining the intraday volatility. A simple definition is given by (before any cross-sectional normalization -- see below)
\begin{eqnarray}\label{hlv}
 &&\beta^{\rm{\scriptstyle{hlv}}}_{is} \equiv  {1\over 2}~\ln\left(U_{is}\right)\\
 &&U_{is} \equiv {1\over d} \sum_{r=1}^d \left({{P^{H}_{i, s+r} - P^L_{i, s+r}}\over P^{C}_{i, s+r}}\right)^2
\end{eqnarray}
Averaging over the last $d$ days is necessary to smooth out the noise. Unlike the ``market beta", however, because we are dealing with a variance-like quantity here (as opposed to a correlation-like quantity), looking back $d$ days is fine here as the high-low intraday swings are much more stable compared with correlation-based betas, which vary dramatically day-to-day with little out-of-sample stability. For our model we use $d=21$ (these are trading days, so this corresponds to 1 month). Also, note that (\ref{hlv}) actually is logarithmic intraday volatility. Taking the intraday volatility itself as a factor beta would have the effect of interfering with the intercept beta (and the resulting t-statistic are worse).

{}Let us mention alternative definitions of $\beta^{\rm{\scriptstyle{hlv}}}_{is}$:
\begin{eqnarray}\label{hlv1}
 &&{\widetilde \beta}^{\rm{\scriptstyle{hlv}}}_{is} \equiv  \ln\left({1\over d} \sum_{r=1}^d {{\left|P^{H}_{i, s+r} - P^L_{i, s+r}\right|}\over P^{C}_{i, s+r}}\right)\\
 &&{\widehat \beta}^{\rm{\scriptstyle{hlv}}}_{is} \equiv  \ln\left({1\over d} \sum_{r=1}^d \ln\left|{P^{C}_{i, s+r} \over P^{O}_{i, s+r}}\right|\right)\label{hlv2}
\end{eqnarray}
Such variations are all fine and have minor pros and cons but do not make it or break it (see below).\footnote{\, Strictly speaking, (\ref{hlv2}) should not be referred to as ``hlv", but we will use this name anyway.} We find that the definition (\ref{hlv}) works rather well.

\subsection{Volume (``Liquidity")}\label{sub.vol}

{}Typically, liquidity is defined (before any cross-sectional normalization -- see below) as a log of the average daily dollar volume (ADDV), which we denote via $D_{is}$:
\begin{equation}\label{ADDV}
 D_{is}\equiv {1\over d} \sum_{r=1}^d V_{i, s+r}~P^C_{i, s+r}
\end{equation}
While the price can change substantially over multiple days $d$, it is the volume $V_{i, s+r}$ that causes the most change in this expression. Let us approximate it as follows:
\begin{eqnarray}
 && D_{is}\approx {\widetilde D}_{is} \equiv P^C_{i,s+1}~{\overline V}_{is}\\
 && {\overline V}_{is} \equiv {1 \over d} \sum_{r=1}^d V_{i, s+r}\label{avg.vol}
\end{eqnarray}
where ${\overline V}_{is}$ is the $d$-day moving average volume. Note that $\ln({\widetilde D}_{is}) = \ln(P^C_{i,s+1}) + \ln({\overline V}_{is})$. However, since we already have the log-of-price based factor beta ({\em i.e.}, prc),\footnote{\, If prc uses $P^{AC}_i$, then vol should also use $P^{AC}_i$ and adjusted volume -- see below.} it is the $\ln({\overline V}_{is})$ piece that could make a difference here. Therefore, we define the volume (``liquidity") factor beta as follows (before any cross-sectional normalization -- see below)
\begin{equation}\label{vol}
 \beta^{\rm{\scriptstyle{vol}}}_{is} \equiv \ln\left({\overline V}_{is}\right)
\end{equation}
Again, averaging over $d$ days is necessary here to smooth out the noise. Note that $V_{i,s+r}$ in (\ref{avg.vol}) is unadjusted, so it jumps during splits. This can be taken care of by adjusting the volume:
\begin{eqnarray}\label{vol1}
 && {\widetilde \beta}^{\rm{\scriptstyle{vol}}}_{is}\equiv \ln\left({\widetilde V}_{is}\right)\\
 && {\widetilde V}_{is} \equiv {1 \over d} \sum_{r=1}^d V_{i, s+r}~{P^C_{i, s+r}\over P^{AC}_{i, s+r}}
\end{eqnarray}
This takes care of the splits, and also inadvertently ``adjusts" the volume for dividends. However, such ``adjustments" are small and do not affect much. The ``in-sample" issue mentioned above as it relates to using adjusted prices is very minor here as it is lost in all the noise in the volume. In regressions below we use $d=21$.

\subsection{Factor Normalization}

{}The factor betas that are expected to have normal distributions, namely, hlv (\ref{hlv}) (also, (\ref{hlv1}) and (\ref{hlv2})) and vol (\ref{vol}) (also, (\ref{vol1})), are then normalized by conforming them to a normal distribution with zero mean and standard deviation equal to the standard deviation of the unnormalized factor beta, as it is done in (Kakushadze and Liew, 2015).\footnote{\, For vol factor one may wish to normalize ADRs separately -- see (Kakushadze and Liew, 2015). In the regressions below we have not done this as this is not critical. Appendix A in (Kakushadze and Liew, 2015) provides R code for normalizing such factors (see {\tt normalize()} function therein).} Factor betas mom (\ref{mom}) and prc (\ref{size}) (also, (\ref{size1})) are not normalized.\footnote{\, Typically, market cap is log-normally distributed (and size normalized); price -- not necessarily.}

\subsection{Universe Selection}\label{sub.univ}

{}Before we can run our regressions, we need to select our universe. Here we are building a factor model for short horizon (overnight) returns. It is clear that such a factor model is not expected to work for stocks that do not trade or trade just a little for days on at a time, {\em i.e.}, our factor model, by its very nature, is expected to work for actively traded, liquid stocks. We wish to keep our discussion here as simple as possible, so we select our universe based on ADDV $D_{is}$ (with $d=21$ -- see (\ref{ADDV})) by taking top 2000 tickers.\footnote{\, One can take top 2500 or some other number -- in practical applications the universe selection is more involved (see below). Above roughly top-3000-by-ADDV the behavior changes -- see below.} However, to ensure that we do not inadvertently introduce a universe selection bias,\footnote{\, {\em I.e.}, to ensure that our results are not a mere consequence of the universe selection.} we do not rebalance the universe daily. Instead, we rebalance monthly, every 21 trading days, to be precise. {\em I.e.}, we break our 5-year backtest period (see below) into 21-day intervals, we compute the universe using ADDV (which, in turn, is computed based on the 21-day period immediately preceding such interval), and use this universe during the entire such interval. The bias that we do have, however, is the survivorship bias. We take the data for the universe of tickers as of 9/6/2014 that have historical pricing data on http://finance.yahoo.com (accessed on 9/6/2014) for the period 8/1/2008 through 9/5/2014. We restrict this universe to include only U.S. listed common stocks and class shares (no OTCs, preferred shares, {\em etc.}) as of 9/6/2014. However, it does not appear that the survivorship bias should be a significant effect. We are not looking at simulated P\&L,\footnote{\, {\em E.g.}, we could trade on the residuals $\varepsilon_{is}$ in (\ref{reg}) ({\em i.e.}, do mean-reversion -- see (Kakushadze, 2015a) for details), which are directly affected by the survivorship bias.} but at the serial t-statistic of the factor returns $f_{As}$ (see below), which are affected by the survivorship bias only indirectly (as $f_{As}$ do not carry the stock index $i$), and on general grounds such second order effects are expected to be lost in all the noise, especially that our regressions are daily, while our ADDV-based universe rebalancing is monthly. Let us also mention that the ADDV-based universe selection is by no means optimal and is chosen here for the sake of simplicity. In practical applications, the factor model should be computed based on the actual trading universe of liquid stocks, which typically is carefully selected based on market cap, liquidity (ADDV), price and other criteria. For such, more optimal universes, the results of the regressions are expected to be even better. However, the ADDV-based universe selection suffices for our purposes here.

\subsection{Regressions}\label{sub2.9}

{}Next, we run the cross-sectional regressions (\ref{reg}) for each date over the period of 5 years.\footnote{\, More precisely, $M = 252\times 5$, and $s=0$ is 9/5/2014 (see the beginning of this section).} This gives the residuals $\varepsilon_{is}$ and factor returns $f_{As}$. We are after the factor return serial t-statistic {\em a la} Fama and MacBeth (1973). These are given in Table \ref{table1} for the top-2000-by-ADDV universe (see previous subsection) for the 4 factors prc, mom, hlv and vol together with int. In Table \ref{table1} we give the annualized t-statistic for the prc factor defined both using adjusted (\ref{size}) and unadjusted (\ref{size1}) close.\footnote{\, We also give median (over the time series) F-statistic. When comparing the F-statistic, the difference in the numbers of factors ({\em e.g.}, int only {\em vs.} int+prc) should be taken into account.}

{}Next, we combine int, prc, mom, hlv and vol and run the regressions on these 5-column factor betas. The results are given in Table \ref{table2} for the top-2000-by-ADDV universe. These results show sizable t-statistic for the 4 factors prc, mom, hlv and vol. We then expand the universe to top-3000-by-ADDV and top-4000-by-ADDV in Table \ref{table3}. The overall median F-statistic drops when we go to top-4000-by-ADDV. The dramatic increase in the t-statistic for mom is due to the fact that open-to-close mean-reversion based on overnight returns on paper appears to ``work" exceptionally well for the lower liquidity stocks with ADDV between top-3000 and top-4000. For these stocks mom has very high t-statistic, and it is also higher for vol,\footnote{\, Thus, in the regression with int+mom+vol for the top-3000-to-4000-by-ADDV universe, we have t-statistic $-9.90$, $-15.72$ and 8.19 for int, mom and vol, respectively.} so these two factors overwhelm prc and hlv in the 4-factor model, albeit the latter have sizable t-statistic on their own ({\em i.e.}, when individually combined with int).

\subsection{Sector Factors}

{}As an additional check, we ran the regressions (\ref{reg}) on a 14-factor model comprised of 10 BICS sectors\footnote{\, Not all tickers in the regressions in Tables \ref{table1}-\ref{table3} have BICS sector data. Those tickers without BICS sector data have been excluded from the regressions reported in Table \ref{table4}, where the top-2000-by-ADDV universe therefore is based on this somewhat smaller subset of tickers.\label{foot.bics}} as factors plus prc, mom, hlv and vol. The results are shown in Table \ref{table4} and further indicate that prc, mom, hlv and vol are statistically relevant predictors for overnight returns. Notice how adding these 4 factors significantly improves sector factor t-statistic (columns 1 and 2 {\em vs.} 3 in Table \ref{table4}). Also, note that the intercept is not added separately as it is already subsumed in the sector betas.\footnote{\, The BICS sector betas are binary: $\beta_{i\alpha} = 1$ if the stock labeled by $i$ belongs to the sector labeled by $\alpha = 1,\dots,10$; otherwise, $\beta_{i\alpha} = 0$. Each stock belongs to one and only one sector. This implies that $\sum_{\alpha=1}^{10} \beta_{i\alpha} = 1$ for each $i$, so a linear combination of sector betas is the intercept.\label{foot.sec}}

\subsection{Intraday Alpha}\label{sub2.11}

{}As one final check that our 4 factors are relevant predictors for overnight returns, we ran a simulation for an intraday mean-reversion alpha in four different incarnations. We take the residuals $\varepsilon_{is}$ of the regression (\ref{reg}), where the factors are i) the intercept only, ii) our 4 factors plus the intercept, iii) BICS sectors only, and iv) our 4 factors plus 10 BICS sectors. The desired {\em dollar} holdings $H_{is}$ are given by
\begin{eqnarray}
 &&H_{is} \equiv -{\widetilde \varepsilon}_{is} ~ {I\over\sum_{j=1}^N \left|{\widetilde \varepsilon}_{js}\right|}\\
 &&\sum_{i=1}^N \left|H_{is}\right| = I\\
 &&\sum_{i=1}^N H_{is} = 0\label{d.n}
\end{eqnarray}
where $I$ is the {\em intraday} investment level, which is the same for all dates $s$. Also, ${\widetilde \varepsilon}_{is}$ are the residuals $\varepsilon_{is}$ in (\ref{reg}) cross-sectionally normalized separately for each date (so everything is out-of-sample) by conforming them to a normal distribution with the same mean and standard deviation as the unnormalized $\varepsilon_{is}$ for such date.\footnote{\, We use the {\tt normalize()} function given in Appendix A of (Kakushadze and Liew, 2015).} Eq. (\ref{d.n}) implies that the portfolio is dollar neutral. This is because $\varepsilon_{is}$ have 0 cross-sectional means (and so do ${\widetilde \varepsilon}_{is}$), which in turn is due to the intercept either being included (cases i)-ii)), or being subsumed in the sector betas (cases iii)-iv)); see footnote \ref{foot.sec}).

{}The portfolio\footnote{\, A similar signal and portfolio can be freely accessed at www.vynance.com.} is established at the open\footnote{\, This is a so-called ``delay-0" alpha -- $P^O_{is}$ is used in the alpha, and as the establishing fill price.} assuming fills at the open prices $P^O_{is}$, and liquidated at the close on the same day assuming fills at the close prices $P^C_{is}$, with no transaction costs or slippage -- our goal here is not to build a trading strategy, but to check if our 4 factors add value.\footnote{\, Hence unweighted regression. Also, here we have the survivorship bias -- see Subsection \ref{sub2.9}.} The P\&L for each stock is
\begin{equation}
 \Pi_{is} = H_{is}\left[{P^C_{is}\over P^O_{is}}-1\right]
\end{equation}
We run our simulation based on the same data as for the regressions in Subsection \ref{sub2.9}, over the same 5-year period and with the same top-2000-by-ADDV universe (as in Table \ref{table4})\footnote{\, This universe is restricted to the tickers with BICS sector data -- see footnote \ref{foot.bics}.} rebalanced every 21 days -- note that, since the alpha is purely intraday, this ``rebalancing" does not generate additional trades, it simply changes the universe that is traded for the next 21 days. The results for the annualized return-on-capital (ROC), annualized Sharpe ratio (SR) and cents-per-share (CPS) are given in Table \ref{table5}. ROC is computed as average daily P\&L divided by the investment level $I$ (with no leverage) and multiplied by 252. SR is computed as daily Sharpe ratio multiplied by $\sqrt{252}$. CPS is computed as the total P\&L divided by total shares traded, with the shares bought plus sold for each stock on each day computed via $Q_{is} = 2 |H_{is}| / P^O_{is}$. Table \ref{table5} and Figure 1 indicate that our 4 factors help improve ROC, SR and CPS.\footnote{\,As a check that this improvement is independent of normalizing the residuals, Table \ref{table6} and Figure 2 give the simulation results for the case where the residuals are not normalized. An evident caveat in this alpha was discussed in (Kakushadze, 2015a), at the end of Section 7.}

\section{Concluding Remarks}\label{sec3}

\noindent
$\bullet$ The prc (``size") factor. In practical applications one will use adjusted yesterday's close data (if available). We also presented the results for rprc as a ``sanity check".\\
\noindent
$\bullet$ The mom (momentum) factor. There is room for improvisation here depending on one's needs. We took an intraday, open-to-close return. One may wish to take other variations, including multiple-day returns. It all depends on one's alpha that the factor model is intended for -- see (Kakushadze and Liew, 2015).\\
$\bullet$ The hlv (intraday volatility) factor. Variations are possible but not game-changing.\\
$\bullet$ The vol (``liquidity") factor. First, in practical applications one will use the average over daily volumes adjusted for splits. It is easier to adjust for both splits and ``dividends" (see Subsection \ref{sub.vol}) if one has fully adjusted close and unadjusted close. The extra adjustment for ``dividends" will not make a material difference. Second, in practical applications one will clean (smooth) outliers in the volume data -- which we purposefully did not do here to keep things simple and transparent and avoid muddying the waters with complexity. Restriction to liquid ({\em e.g.}, top-2000-by-ADDV) stocks provides indirect outlier ``management"; however, in practical applications one can and should deal with outliers directly as this will improve the results.\\
$\bullet$ Universe selection. First, for the reasons outlined in Subsection \ref{sub.univ}, we do not expect the survivorship bias to be a significant effect for the t-statistic of the factor returns. Second, to keep things simple and transparent, we based our universe on ADDV (top-2000-by-ADDV). In practical applications one will compute the factors based on a typical trading universe of liquid stocks, which is selected based on market cap, liquidity (ADDV), price, {\em etc.} This will improve the results further.\\
$\bullet$ Sectors/Industries. Here we also added 10 BICS sectors to our 4-factor model, which (not surprisingly) improves the statistic. In practical applications instead of sectors one will use a more granular level, {\em e.g.}, in the particular example of BICS we have the hierarchy ``sector $\rightarrow$ industry $\rightarrow$ sub-industry", so one will go down to the sub-industry level. To avoid overly granular sub-industries ({\em i.e.}, sub-industries with too few stocks), one can require a minimum of, say, 10 stocks per sub-industry and then small sub-industries are pruned to the industry level and, if need be, to the sector level. This way one obtains many more (pruned) sub-industries (than the number of sectors) and the resulting statistic is expected to be even better once these sub-industries are used as factors together with prc, mom, hlv and vol. One can take this a step further and build a full-fledged multi-factor risk model by computing the factor covariance matrix and specific risk, which is a proprietary topic outside of the scope of this note.\\
$\bullet$ Market beta. For the reasons mentioned in Subsection \ref{sub.beta}, we simply used the intercept in lieu of any market beta. One can compute the market beta in the standard way based on stock correlations with the market factor. Alternatively, one can take the first principal component of the stock sample covariance matrix. Either way, due to the inherent out-of-sample instability of the correlation matrix, such beta factors are not expected to be particularly stable out-of-sample.\\
$\bullet$ Intraday. Our 4-factor model is expected to work reasonably well beyond overnight returns, {\em i.e.}, intraday. Not necessarily on microsecond time horizons, but for, {\em e.g.}, intraday strategies that establish around the open and liquidate before the close.\\
$\bullet$ Implied volatility. When commercial risk model providers pitch their products, one argument is that using option implied volatility (which, to start with, is available only for optionable stocks, and this is already an important limitation) to model stock volatility should work better,\footnote{\, In this context, the (apparently inconclusive) paper (Ederington and Guan, 2002) sometimes is referred to.} and if a portfolio manager does not possess the implied volatility data or the know-how for incorporating it into a risk model organically, he or she would be better off simply buying a risk model from a provider. However, as was recently pointed out in (Kakushadze, 2015b), this argument appears to be thin, at best. Nowadays, with ever-shortening lookbacks, it is unclear if the implied volatility indeed adds any value when the risk model is used in actual portfolio optimization (or regression) for actual alphas. In this regard a new study, which is outside of the scope of this paper, would appear to be warranted.
\\
\\
\noindent
{}{\bf Acknowledgments.} I would like to thank Jim Liew for valuable discussions.
\\
\\
\noindent
{}{\bf Legalese.} The risk factors discussed in this paper are used in a variant of the source code for Quantigic$^\circledR$ Risk Model$^{\rm{\scriptstyle{TM}}}$ and are provided herein with the express permission of Quantigic$^\circledR$ Solutions LLC. The copyright owner of said source code retains all rights, title and interest therein and any and all copyrights therefor.



\begin{table}[ht]
\caption{Results for regressions (\ref{reg}) with $\beta_{iAs}$ for each date $s$ consisting of one column (int only) or two columns (int+X); int = intercept; X = prc (\ref{size}), rprc (\ref{size1}), mom (\ref{mom}), hlv (\ref{hlv}) or vol (\ref{vol}); prc = log(adjusted previous close); rprc = log(unadjusted previous close). t-stat:$\star$ means annualized (via multiplying daily t-statistic by $\sqrt{252}$) t-statistic for the corresponding factor $\star$ {\em a la} Fama and MacBeth (1973). We also give the median F-statistic (over the time series of F-statistic for daily cross-sectional regressions). For comparative purposes we also give the corresponding results for alternative definitions of hlv (hlv1 (\ref{hlv1}) and hlv2 (\ref{hlv2})) and vol1 (\ref{vol1}). The universe is top-2000-by-ADDV rebalanced every 21 days.} 
\begin{tabular}{l l l l} 
\\
\hline\hline 
Regression & F-stat & t-stat:int & t-stat:X\\ [0.5ex] 
\hline 
int only & 243.8 & 1.09 & --- \\ 
int+prc & 127.4 & 2.65 & -6.02\\
int+rprc & 126.8 & 2.52 & -5.29\\
int+mom & 127.6 & 1.00 & -2.82\\
int+hlv & 133.8 & 2.88 & 3.64\\
int+vol & 128.2 & -2.45 & 3.83\\
int+hlv1 & 134.5 & 2.91 & 3.67\\
int+hlv2 & 134.4 & 2.90 & 3.55\\
int+vol1 & 127.6 & -2.09 & 4.11 \\ [1ex] 
\hline 
\end{tabular}
\label{table1} 
\end{table}

\begin{table}[ht]
\caption{Results for regressions (\ref{reg}) with int plus 4 factors prc, mom, hlv and vol. The notations are the same as in Table \ref{table1}. The universe is top-2000-by-ADDV rebalanced every 21 days.} 
\begin{tabular}{l l l l l l } 
\\
\hline\hline 
t-stat:int & t-stat:prc & t-stat:mom & t-stat:hlv & t-stat:vol & F-stat\\
2.28 & -4.66 & -3.78 & 2.95 & 2.91 & 56.6\\ [0.5ex] 
\hline 
t-stat:int & t-stat:rprc & t-stat:mom & t-stat:hlv & t-stat:vol & F-stat\\
2.19 & -3.84 & -3.79 & 2.99 & 3.00 & 56.5\\
\hline 
t-stat:int & t-stat:prc & t-stat:mom & t-stat:hlv & t-stat:vol1 & F-stat\\
2.25 & -4.43 & -3.77 & 2.96 & 3.17 & 56.5\\
\hline 
t-stat:int & t-stat:rprc & t-stat:mom & t-stat:hlv & t-stat:vol1 & F-stat\\
2.10 & -4.01 & -3.78 & 3.00 & 3.52 & 56.5\\ [1ex] 
\hline 
\end{tabular}
\label{table2} 
\end{table}

\begin{table}[ht]
\caption{Results for regressions (\ref{reg}) with int plus 4 factors prc, mom, hlv and vol. The notations are the same as in Table \ref{table1}. The universe is top-3000-by-ADDV and top-4000-by-ADDV (as indicated) rebalanced every 21 days.} 
\begin{tabular}{l l l l l l l} 
\\
\hline\hline 
Univ & t-stat:int & t-stat:prc & t-stat:mom & t-stat:hlv & t-stat:vol & F-stat\\
3000 & 2.82 & -5.92 & -4.80 & 2.85 & 2.76 & 70.2\\ [0.5ex] 
\hline 
Univ & t-stat:int & t-stat:rprc & t-stat:mom & t-stat:hlv & t-stat:vol & F-stat\\
3000 & 2.72 & -5.91 & -4.81 & 2.79 & 2.71 & 70.2\\
\hline 
Univ & t-stat:int & t-stat:prc & t-stat:mom & t-stat:hlv & t-stat:vol & F-stat\\
4000 & 2.12 & -3.71 & -12.96 & 2.60 & 3.54 & 49.9\\
\hline 
Univ & t-stat:int & t-stat:rprc & t-stat:mom & t-stat:hlv & t-stat:vol & F-stat\\
4000 & 1.56 & -4.83 & -12.91 & 2.06 & 3.80 & 49.7\\ [1ex] 
\hline 
\end{tabular}
\label{table3} 
\end{table}
%
\begin{table}[ht]
\caption{Results for regressions (\ref{reg}) with the 10 BICS sectors (as factors) with (the first two columns) and without (the last column) the 4 factors prc, mom, hlv and vol. S = 10 BICS sectors labeled by S1,\dots, S10. The rest of the notations are the same as in Table \ref{table1}. The universe is top-2000-by-ADDV rebalanced every 21 days. When comparing F-statistic, the difference in the numbers of factors (14 in the first two columns {\em vs.} 10 in the last column) should be taken into account, so adding the 4 factors actually improves F-statistic.} 
\begin{tabular}{l l l l} 
\\
\hline\hline 
Factor/Regression & S+prc+mom+hlv+vol & S+rprc+mom+hlv+vol & S only\\[0.5ex] 
\hline 
F-stat & 22.7 & 22.7 & 28.4\\ 
t-stat:prc & -4.83 & --- & ---\\
t-stat:rprc & --- & -4.12 & ---\\
t-stat:mom & -4.35 & -4.36 & ---\\
t-stat:hlv & 3.16 & 3.26 & ---\\
t-stat:vol & 3.12 & 3.23 & ---\\
t-stat:S1 & 2.51 & 2.45 & 0.84\\
t-stat:S2 & 2.33 & 2.26 & 1.08\\
t-stat:S3 & 2.51 & 2.44 & 1.43\\
t-stat:S4 & 2.51 & 2.44 & 1.58\\
t-stat:S5 & 2.11 & 2.05 & 0.70\\
t-stat:S6 & 2.33 & 2.28 & 0.90\\
t-stat:S7 & 2.33 & 2.26 & 0.91 \\
t-stat:S8 & 2.33 & 2.27 & 1.24 \\
t-stat:S9 & 2.35 & 2.30 & 1.09\\
t-stat:S10 & 2.70 & 2.66 & 1.14\\ [1ex] 
\hline 
\end{tabular}
\label{table4} 
\end{table}

\begin{table}[ht]
\caption{Simulation results for the intraday mean-reversion alphas discussed in Subsection \ref{sub2.11}. The notations are the same as in Table \ref{table1} and Table \ref{table4}. The universe is top-2000-by-ADDV rebalanced every 21 days.} 
\begin{tabular}{l l l l} 
\\
\hline\hline 
Model & ROC & SR & CPS\\[0.5ex] 
\hline 
int only & 26.94\% & 7.79 & 0.82\\
int+prc+mom+hlv+vol & 28.79\% & 11.22 & 0.87\\
int+rprc+mom+hlv+vol & 28.80\% & 11.23 & 0.87\\
S only & 33.27\% & 11.55 & 1.02\\
S+prc+mom+hlv+vol & 34.30\% & 14.94 & 1.04\\
S+rprc+mom+hlv+vol & 34.34\% & 14.97 & 1.04\\ [1ex] 
\hline 
\end{tabular}
\label{table5} 
\end{table}

\begin{table}[ht]
\caption{Simulation results for the intraday mean-reversion alphas discussed in Subsection \ref{sub2.11}, but without normalizing the residuals. The notations are the same as in Table \ref{table1} and Table \ref{table4}. The universe is top-2000-by-ADDV rebalanced every 21 days.} 
\begin{tabular}{l l l l} 
\\
\hline\hline 
Model & ROC & SR & CPS\\[0.5ex] 
\hline 
int only & 38.64\% & 5.14 & 1.01\\
int+prc+mom+hlv+vol & 40.24\% & 5.84 & 1.04\\
int+rprc+mom+hlv+vol & 40.26\% & 5.85 & 1.04\\
S only & 44.58\% & 6.21 & 1.17\\
S+prc+mom+hlv+vol & 45.29\% & 6.80 & 1.18\\
S+rprc+mom+hlv+vol & 45.32\% & 6.81 & 1.18\\ [1ex] 
\hline 
\end{tabular}
\label{table6} 
\end{table}

\begin{figure}[ht]
\centerline{\epsfxsize 4.truein \epsfysize 4.truein\epsfbox{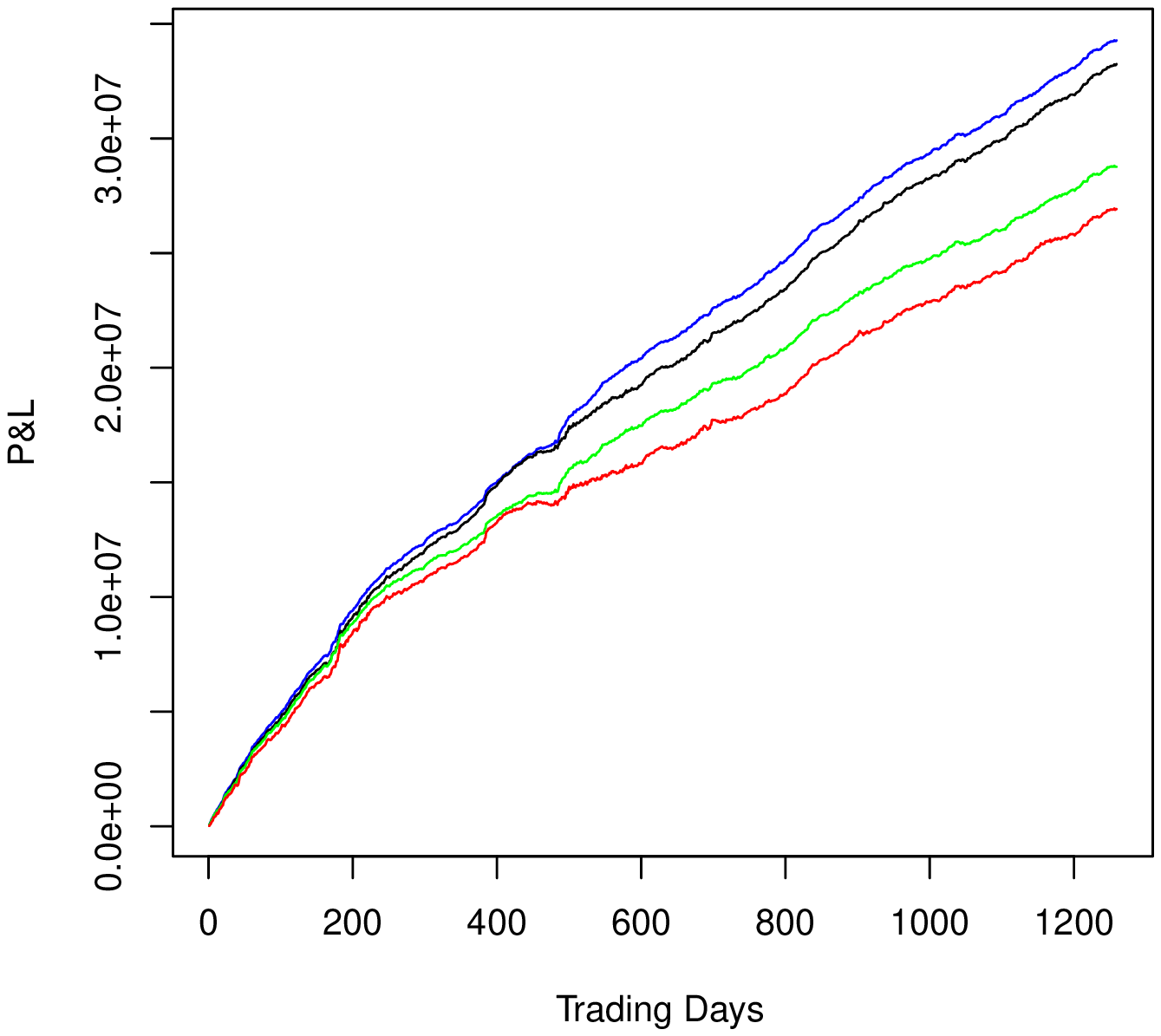}}
\noindent{\small {Figure 1. P\&L graphs for the mean-reversion alpha discussed in Subsection \ref{sub2.11}, with a summary in Table \ref{table5}. Bottom-to-top-performing: i) the intercept only, ii) our 4-factors plus the intercept, iii) 10 BICS sectors only, and iv) our 4-factors plus 10 BICS sectors. The investment level is \$10M long plus \$10M short.}}
\end{figure}

\begin{figure}[ht]
\centerline{\epsfxsize 4.truein \epsfysize 4.truein\epsfbox{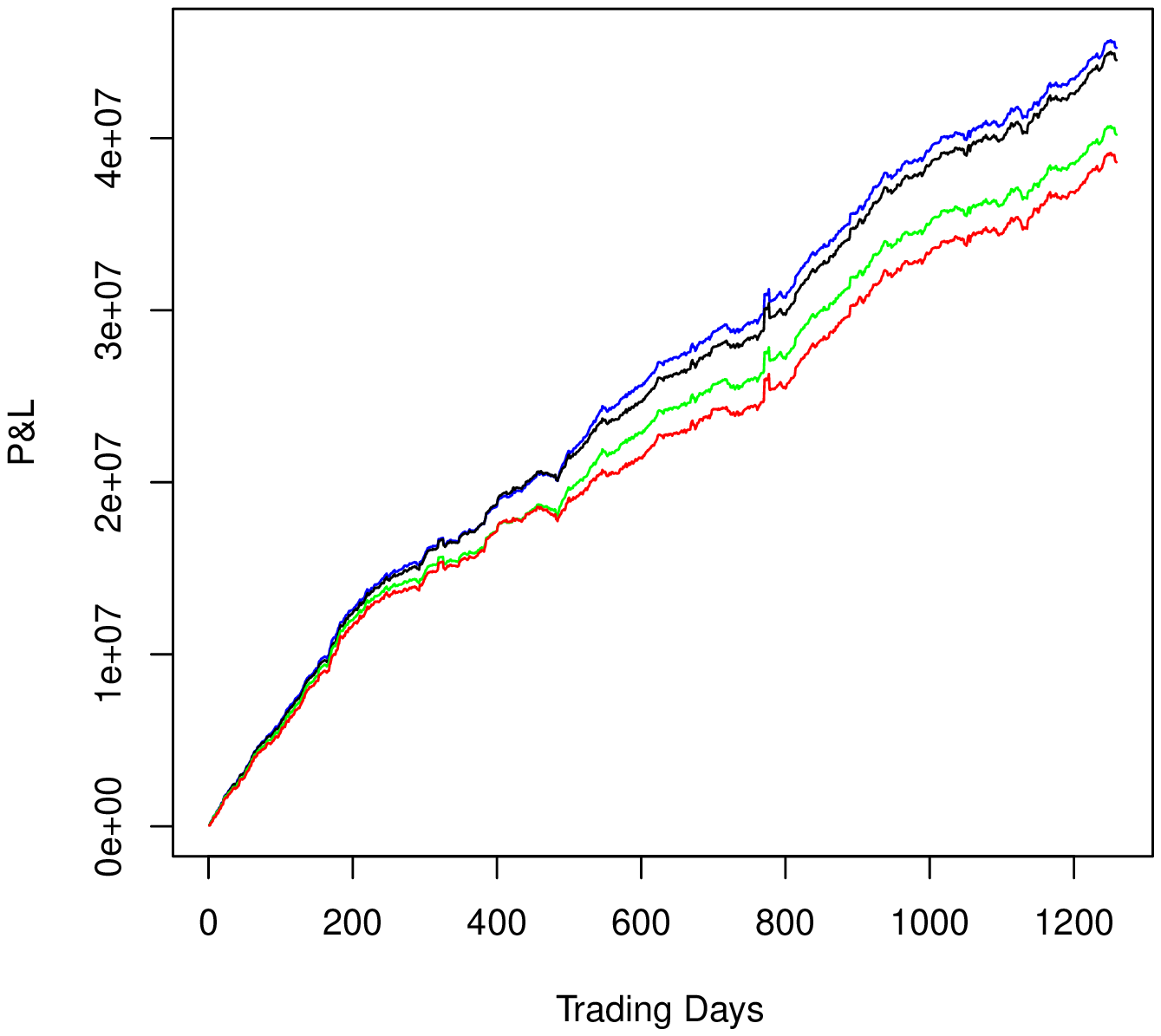}}
\noindent{\small {Figure 2. P\&L graphs for the mean-reversion alpha discussed in Subsection \ref{sub2.11}, but without normalizing the regression residuals, with a summary in Table \ref{table6}. Bottom-to-top-performing: i) the intercept only, ii) our 4-factors plus the intercept, iii) 10 BICS sectors only, and iv) our 4-factors plus 10 BICS sectors. The investment level is \$10M long plus \$10M short.}}
\end{figure}

\end{document}